\documentclass[proceedings]{JHEP3}
\usepackage{amsfonts}
\usepackage{amsmath}
\usepackage{epsfig,multicol}

\newbox\mybox

\newcommand\fverb{\setbox\mybox=\hbox\bgroup\verb}
\newcommand\fverbdo{\egroup\medskip\noindent\fbox{\unhbox\mybox}\ }
\newcommand\fverbit{\egroup\item[\fbox{\unhbox\mybox}]}
\conference{$\mathcal{PT}$-symmetric deformations of Calogero models}
\abstract{We demonstrate that Coxeter groups allow for complex $\mathcal{PT}$-symmetric deformations across the boundaries of all Weyl chambers. We compute the explicit deformations for the $A_2$ and $G_2$-Coxeter group and apply these constructions
to Calogero-Moser-Sutherland models invariant under the extended Coxeter groups. The eigenspecta for the deformed models are real and contain the spectra of the undeformed case as subsystem.}

\title{$\mathcal{PT}$-symmetric deformations of Calogero models}
\author{Andreas Fring$^\bullet$ and Miloslav Znojil$^\circ$ \\
$^\bullet$ Centre for Mathematical Science, City University, \\
$\;$ Northampton Square, London EC1V 0HB, UK\\
$^\circ$ Nuclear Physics Institute ASCR, 250 68\ \v{R}e\v{z}, Czech
Republic \\
$\;$ E-mail: a.fring@city.ac.uk , znojil@ujf.cas.cz}

\input{tcilatex}

\begin{document}

\section{Introduction}

The study of pseudo-Hermitian Hamiltonian systems has attracted a
considerable amount of attention in the last few years. For recent reviews
and special issues devoted to this topic see \cite%
{special,specialCzech,Benderrev,special2}. One of the main reasons for the
popularity of these types of Hamiltonians is the fact that they possess real
eigenvalue spectra, despite of being non-Hermitian, and therefore constitute
interesting candidates for a new sort of stable physical systems overlooked
up to now. Alternatively to using the concept of pseudo-Hermiticity \cite%
{Mostafazadeh:2002hb} or quasi-Hermiticity \cite{Dieu,Will} one may
equivalently explain the reality of the spectrum of some non-Hermitian
Hamiltonians when one encounters unbroken $\mathcal{PT}$-symmetry, which in
the recent context was first pointed out in \cite{Bender:2002vv}. Unbroken
specifies that both the Hamiltonian \textit{and} the wavefunction remain
invariant under a simultaneous parity transformation $\mathcal{P}%
:x\rightarrow -x$ and time reversal $\mathcal{T}:t\rightarrow -t$. When
acting on complex valued functions the anti-linear operator $\mathcal{T}$ is
understood to act as complex conjugation.

These observations can be exploited in the construction of new interesting
models with real eigenvalue spectra when taking previously studied Hermitian
examples as starting points. The above statements imply that one has two
possibilities at hand. One could either employ pseudo-Hermiticity, which
involves the usually technically difficult task to construct a meaningful
metric, e.g. \cite{JM,CA,KBZ,Mostdel,MGH,PEGA}, or in contrast use $\mathcal{%
PT}$-symmetry as a very transparent and simple principle, at least on the
level of the Hamiltonian itself. Starting with a $\mathcal{PT}$-symmetric
Hamiltonian or less restrictive with a parity invariant potential system one
may extend such type of models by adding $\mathcal{PT}$-symmetric terms to
it or by deforming existing terms in a $\mathcal{PT}$-symmetric manner. The
latter construction principle has been applied to a huge number of models,
notably the harmonic oscillator in \cite{Bender:1998ke}, which constitutes
the starting point of the current activities in this field of research.

In the context of Calogero models \cite{Cal1,Cal2,Cal3} such type of
extensions were first carried out in \cite%
{Basu-Mallick:2000af,Basu-Mallick:2001ce} by adding $\mathcal{PT}$-symmetric
terms to the $A_{n}$ and $B_{n}$-Calogero models. Shortly afterwards an
alternative procedure was proposed in \cite{Milos}, where the $A_{2}$%
-Calogero model was genuinely deformed in a $\mathcal{PT}$-symmetric manner.
The analysis in \cite{Basu-Mallick:2000af} was extended thereafter in \cite%
{AF} to Calogero models related to all Coxeter groups and also generalized
to the larger class of Calogero-Moser-Sutherland (CMS) models \cite%
{Cal1,Cal2,Cal3,Suth3,Suth4,Mo,OP2,Per} involving more general types of
potentials rather than the rational one. Other versions of deformations of
CMS-models have also been proposed for instance in \cite{FKq}, albeit a
concrete relation to $\mathcal{PT}$-symmetry had not been established, even
though it is easy to verify that the models constructed in \cite{FKq} are
also $\mathcal{PT}$-symmetric. The purpose of this paper is to provide the
general mathematical framework for the deformation carried out in \cite%
{Milos} and generalize the construction to all Coxeter groups and more
general potentials. Thereafter we study some of the physical properties of
the newly obtained models.

Our manuscript is organized as follows: In order to fix our conventions we
recall in section 2 some of the basic features of CMS-models and indicate
the structure we expect to find for the deformed models. In section 3 we
demonstrate how Coxeter groups may be systematically deformed in a $\mathcal{%
PT}$-symmetric manner. We illustrate the general setting with the two
explicit examples of the $A_{2}$ and $G_{2}$-Coxeter group. We apply these
construction in section 4 to CMS-models, which are invariant under the
extended Coxeter group. We show that models for which this invariance is
broken in a particular way also possess interesting properties. Thereafter
we specialize to the Calogero models and construct their eigensystems for
some specific deformations. The key finding is that some constraints on the
parameter space of the model resulting from physical requirements may be
relaxed in the deformed model. For some simple extended model we demonstrate
that the energy spectrum is real and contains the one of the undeformed case
as a subsystem. We state our conclusions in section 5.

\section{Extended symmetries for Calogero-Moser-Sutherland models}

Let us briefly recall some features of the CMS-models, which will be
relevant for our analysis. The models describe $n$ particles moving on a
line, whose coordinates $q$ and canonically conjugate momenta $p$ may be
assembled into vectors $q,p\in \mathbb{R}^{n}$. The Hamiltonian for the
CMS-models related to all Coxeter groups $\mathcal{W}$ may be written
generically as 
\begin{equation}
\mathcal{H}_{\text{CMS}}=\frac{p^{2}}{2}+\frac{m^{2}}{16}\sum_{\alpha \in
\Delta _{s}}(\alpha \cdot q)^{2}+\frac{1}{2}\sum\limits_{\alpha \in \Delta
}g_{\alpha }V(\alpha \cdot q)\qquad m,g_{\alpha }\in \mathbb{R}.  \label{Ham}
\end{equation}
The dimensionality of the space in which the roots $\alpha $ of the root
system $\Delta $ are realized is $n$. The sum in the confining term of the
potential only extends over the short roots $\Delta _{s}$. One may impose
further restrictions on the coupling constants $g_{\alpha }$ in order to
guarantee integrability \cite{OP2,Per,FM} and invariance of the Hamiltonian
under the action of $\mathcal{W}$. The latter demands that $g_{\alpha
}=g_{\beta }$ when the roots $\alpha $ and $\beta $ have the same length,
i.e. if $\alpha ^{2}=\beta ^{2}$. When the potential $V$ is taken to be $%
V(x)=1/x^{2}$ the Hamiltonian (\ref{Ham}) constitutes the Calogero model,
whereas the generalized CMS-models are obtained by choosing $V(x)=1/\sin
^{2}x$, $V(x)=1/\sinh ^{2}x$ or $V(x)=1/\func{sn}^{2}x$.

A key feature of the model (\ref{Ham}) for our purposes is that it admits
the entire Coxeter group $\mathcal{W}$ as a symmetry, i.e. 
\begin{eqnarray}
\mathcal{H}_{\text{CMS}} &=&\frac{\sigma _{i}p\cdot \sigma _{i}p}{2}+\frac{%
m^{2}}{16}\sum_{\alpha \in \Delta _{s}}(\alpha \cdot \sigma _{i}q)^{2}+\frac{%
1}{2}\sum\limits_{\alpha \in \Delta }g_{\alpha }V(\alpha \cdot \sigma _{i}q),
\label{inv} \\
&=&\frac{p^{2}}{2}+\frac{m^{2}}{16}\sum_{\alpha \in \Delta _{s}}(\sigma
_{i}^{-1}\alpha \cdot q)^{2}+\frac{1}{2}\sum\limits_{\alpha \in \Delta
}g_{\alpha }V(\sigma _{i}^{-1}\alpha \cdot q)  \notag
\end{eqnarray}
where $\sigma _{i}$ can be any Weyl reflection (\ref{Weyl}). For the
confining term to remain invariant we need to use that short roots are
mapped into short root by the entire Coxeter group. This symmetry stipulates
that these models are invariant with respect to various parity
transformations $\mathcal{P}$ across the hyperplanes through the origin
orthogonal to the root $\alpha _{i}$ or in other words across the boundaries
of all Weyl chambers.

Our aim is here to modify the models such that they remain invariant under
the action of the newly defined $\mathcal{PT}$-symmetrically extended
Coxeter group, which we denote by $\mathcal{W}^{\mathcal{PT}}$. We propose
the new Hamiltonians to be of the form 
\begin{equation}
\mathcal{H}_{\mathcal{PT}\text{CMS}}=\frac{p^{2}}{2}+\frac{m^{2}}{16}\sum_{%
\tilde{\alpha}\in \tilde{\Delta}_{s}}(\tilde{\alpha}\cdot q)^{2}+\frac{1}{2}%
\sum\limits_{\tilde{\alpha}\in \tilde{\Delta}}g_{\tilde{\alpha}}V(\tilde{%
\alpha}\cdot q)\qquad m,g_{\tilde{\alpha}}\in \mathbb{R},  \label{PTCM}
\end{equation}%
where we have replaced the standard roots $\alpha \in \Delta $ by deformed
roots $\tilde{\alpha}\in \tilde{\Delta}$. Formally $\mathcal{H}_{\text{CMS}}$
and $\mathcal{H}_{\mathcal{PT}\text{CMS}}$ are very similar, with the
crucial difference that the latter is in general complex and non-Hermitian.

Nonetheless, the $\mathcal{PT}$-symmetry can be utilised to establish the
reality of the spectrum with a minor modification. As we have complexified
here each Weyl reflection across any hyperplane orthogonal to every root we
have as many $\mathcal{PT}$-operators, i.e. anti-linear operators, as
hyperplanes. This means we can employ any of these operators in the standard
argument\footnote{%
By construction $\tilde{\sigma}_{\alpha _{i}}$, see (\ref{sdef}) for
definition, is a symmetry of the new Hamiltonian $\mathcal{H}_{\mathcal{PT}%
\text{CMS}}$, that is we have $[\mathcal{H}_{\mathcal{PT}\text{CMS}},\tilde{%
\sigma}_{\alpha _{i}}]=0$. Assuming further that the eigenfunctions are also
invariant with regard to $\mathcal{W}^{\mathcal{PT}}$, i.e. $\tilde{\sigma}%
_{\alpha _{i}}\Phi =\Phi $, the reality of the eigenspectrum follows
trivially from 
\begin{equation*}
\varepsilon \Phi =\mathcal{H}_{\mathcal{PT}\text{CMS}}\Phi =\mathcal{H}_{%
\mathcal{PT}\text{CMS}}\tilde{\sigma}_{\alpha _{i}}\Phi =\tilde{\sigma}%
_{\alpha _{i}}\mathcal{H}_{\mathcal{PT}\text{CMS}}\Phi =\tilde{\sigma}%
_{\alpha _{i}}\mathcal{\varepsilon }\Phi =\mathcal{\varepsilon }^{\ast }%
\tilde{\sigma}_{\alpha _{i}}\Phi =\mathcal{\varepsilon }^{\ast }\Phi .
\end{equation*}%
}. In turn this also means that we could in principle make our construction
less constraining by demanding less symmetry. What is of course not known at
this point is whether the wavefunctions of the deformed Hamiltonian also
respect the extended symmetry. However, as we shall demonstrate below with
some concrete examples this will indeed be the case.

In order to see that such type of models really exist and how these models
can be constructed we need to assemble first some mathematical tools and
establish the fact that one can indeed construct a meaningful set of
deformed roots $\tilde{\alpha}$.

\section{$\mathcal{PT}$-symmetric deformations of Coxeter groups}

We recall, see e.g. \cite{Hum,HUm2}, that a Coxeter group $\mathcal{W}$ is
generated by the Weyl reflections $\sigma _{i}$ associated with a set of
simple roots $\{\alpha _{i}\}$ which span the entire root space $\Delta $%
\begin{equation}
\sigma _{i}(x)=x-2\frac{x\cdot \alpha _{i}}{\alpha _{i}^{2}}\alpha
_{i}\qquad \text{for }1\leq i\leq \ell \equiv \text{rank }\mathcal{W}\text{; 
}x,\alpha _{i}\in \mathbb{R}^{n}.  \label{Weyl}
\end{equation}%
The roots may be represented in various different Euclidean spaces with
dimensionality not necessarily equal to $\ell $. Here our aim is to
construct a complex extended root system $\tilde{\Delta}(\varepsilon )$
containing the roots $\tilde{\alpha}_{i}(\varepsilon )$ represented in $%
\mathbb{R}^{n}\oplus i\mathbb{R}^{n}$, depending on some deformation
parameter $\varepsilon \in \mathbb{R}$. We demand that each deformed root
reduces one-to-one to a root in the root space $\Delta $ 
\begin{equation}
\lim_{\varepsilon \rightarrow 0}\tilde{\alpha}_{i}(\varepsilon )=\alpha
_{i}\qquad \text{for }\tilde{\alpha}_{i}(\varepsilon )\in \tilde{\Delta}%
(\varepsilon ),\alpha _{i}\in \Delta ,  \label{def}
\end{equation}%
such that the entire root space reduces as 
\begin{equation}
\lim_{\varepsilon \rightarrow 0}\tilde{\Delta}(\varepsilon )=\Delta .
\end{equation}%
Furthermore, we require that the extended root system $\tilde{\Delta}%
(\varepsilon )$ remains invariant under the $\mathcal{PT}$-symmetrically
extended Coxeter group $\mathcal{W}^{\mathcal{PT}}$. Note that in principle
we may choose any of the hyperplanes through the origin orthogonal to a root 
$\alpha _{i}\in \Delta $ across which the parity symmetry $\mathcal{P}$ can
be extended to a $\mathcal{PT}$-symmetry. Thus we could expect $\ell h\cdot
\ell h/2$ deformed roots, with $h$ denoting the Coxeter number and $\ell h$
being the total number of roots. However, the deformations to any of the
hyperplanes can in fact be made equivalent and the replacement 
\begin{equation}
\alpha _{i}\rightarrow \tilde{\alpha}_{i}(\varepsilon )\qquad \text{for }%
1\leq i\leq \ell h,
\end{equation}%
becomes indeed one-to-one as we shall see below. \noindent

From these requirements we may now attempt to construct the root system $%
\tilde{\Delta}(\varepsilon )$. We start by selecting a particular root $%
\alpha _{i}$, which does not have to be simple, and perform a complex $%
\mathcal{PT}$-symmetric extension across the hyperplane through the origin
orthogonal to this root. This deformation leads to a new, so far unspecified
root $\tilde{\alpha}_{i}(\varepsilon )$. Studying now the properties of this
root will enable us to determine it. Decomposing the complex extended Weyl
reflection into a product of standard Weyl reflections (\ref{Weyl}) and a
complex conjugation (time-reversal) as 
\begin{equation}
\tilde{\sigma}_{\alpha _{i}}:=\sigma _{\alpha _{i}}\mathcal{T},  \label{sdef}
\end{equation}%
we compute its action on a root 
\begin{eqnarray}
\tilde{\sigma}_{\alpha _{j}}\left( \tilde{\alpha}_{j}(\varepsilon )\right)
&=&\sigma _{\alpha _{j}}\mathcal{T}\left( \func{Re}\tilde{\alpha}%
_{j}(\varepsilon )\right) +\sigma _{\alpha _{j}}\mathcal{T}\left( i\func{Im}%
\tilde{\alpha}_{j}(\varepsilon )\right) \\
&=&\sigma _{\alpha _{j}}\left( \func{Re}\tilde{\alpha}_{j}(\varepsilon
)\right) -i\sigma _{\alpha _{j}}\left( \func{Im}\tilde{\alpha}%
_{j}(\varepsilon )\right) \\
&=&-\func{Re}\tilde{\alpha}_{j}(\varepsilon )-i\func{Im}\tilde{\alpha}%
_{j}(\varepsilon )  \label{xx} \\
&=&-\tilde{\alpha}_{j}(\varepsilon ).  \label{x}
\end{eqnarray}%
In view of (\ref{def}) we demanded here that the complex extended Weyl
reflection $\tilde{\sigma}_{\alpha _{i}}$ maps the deformed root $\tilde{%
\alpha}_{i}(\varepsilon )$ into its negative which should in view of the
limit (\ref{def}) also hold for the real part independently. For the
remaining term of the root the minus sign is created by the complex
conjugation $\mathcal{T}$, such the imaginary part has to be invariant under
the Weyl reflection, i.e. it has to be a vector lying in the hyperplane
across which the reflection is carried out. Comparing now (\ref{xx}) and (%
\ref{x}) we find as solution for $\tilde{\alpha}_{i}(\varepsilon )$%
\begin{equation}
\func{Re}\tilde{\alpha}_{i}(\varepsilon )=R(\varepsilon )\alpha _{i}\text{%
\qquad and\qquad }\func{Im}\tilde{\alpha}_{i}(\varepsilon )=I(\varepsilon
)\sum\nolimits_{j\neq i}\kappa _{j}\lambda _{j},  \label{hj}
\end{equation}%
where $\kappa _{j}\in \mathbb{R}$ and the $\lambda _{i}$ have to be elements
of the weight lattice, i.e. they are orthogonal to the simple roots $%
2\lambda _{i}\cdot $ $\alpha _{j}/\alpha _{j}^{2}=\delta _{ij}$. The real
valued functions $R(\varepsilon )$ and $I(\varepsilon )$ are arbitrary at
this stage, with the only condition to satisfy 
\begin{equation}
\lim_{\varepsilon \rightarrow 0}R(\varepsilon )=1\text{\qquad and\qquad }%
\lim_{\varepsilon \rightarrow 0}I(\varepsilon )=0,  \label{RI}
\end{equation}%
in order to fulfill the requirement (\ref{def}). Note that $R(\varepsilon )$
and $I(\varepsilon )$ may also be multiplied by any invariant of the
extended Weyl group $\mathcal{W}^{\mathcal{PT}}$.

The remaining roots can be constructed by acting with all possible
non-equivalent $\ell h-1$ reflections $\sigma _{\alpha _{i}}\mathcal{T}$ \
on these roots and hence producing the anticipated number of $\ell h$ roots $%
\tilde{\alpha}_{i}(\varepsilon )$ $\in \tilde{\Delta}(\varepsilon )$.
Supposing now we have constructed a new root as $\beta =\tilde{\sigma}%
_{\alpha _{k}}(\tilde{\alpha}_{i})$, it is then clear that by construction
also for that new root the imaginary part is orthogonal to its real part 
\begin{equation}
\func{Re}\beta \cdot \func{Im}\beta =\sigma _{\alpha _{k}}(\func{Re}\tilde{%
\alpha}_{i})\cdot \sigma _{\alpha _{k}}(\func{Im}\tilde{\alpha}_{i})=\func{Re%
}\tilde{\alpha}_{i}\cdot \func{Im}\tilde{\alpha}_{i}=0.
\end{equation}%
The property which is, however, not guaranteed is that the decomposition of
the undeformed root into a sum over simple roots $\alpha =\sum_{i=1}\alpha
_{i}$ is preserved by the deformation. Nonetheless, we shall verify this
feature for the explicit examples below. Sometimes we can even find 
\begin{equation}
\tilde{\alpha}_{i}\cdot \tilde{\alpha}_{j}=\tilde{\sigma}_{\alpha _{k}}(%
\tilde{\alpha}_{i})\cdot \tilde{\sigma}_{\alpha _{k}}(\tilde{\alpha}_{j}),
\label{sd}
\end{equation}%
for which there is also no general justification. When (\ref{sd}) holds we
can even impose a stronger constraint and require that inner products of
roots and deformed roots are identical 
\begin{equation}
\alpha _{i}\cdot \alpha _{j}=\tilde{\alpha}_{i}\cdot \tilde{\alpha}_{j},
\end{equation}%
which allows us to fix the functions $R(\varepsilon )$ and $I(\varepsilon )$.

Alternatively to the above construction we may also deform each root as 
\begin{equation}
\alpha _{i}\rightarrow \tilde{\alpha}_{i}(\varepsilon )=R(\varepsilon
)\alpha _{i}\pm iI(\varepsilon )\alpha _{i}\qquad \text{for }\alpha _{i}\in
\Delta _{\pm }.  \label{def2}
\end{equation}%
Note that in this deformation positive and negative roots in $\tilde{\Delta}%
_{+}$ and $\tilde{\Delta}_{-}$ are no longer related by an overall minus
sign as in $\Delta _{+}$ and $\Delta _{-}$, where $\alpha _{i}\in \Delta
_{+} $ always has a counter part $-\alpha _{i}\in \Delta _{-}$. However, by
construction we still have the property 
\begin{equation}
\tilde{\sigma}_{\alpha _{i}}\left( \tilde{\alpha}_{i}(\varepsilon )\right) =-%
\tilde{\alpha}_{i}(\varepsilon ),  \label{alt}
\end{equation}%
which is needed to achieve invariance under the extended Coxeter group $%
\mathcal{W}^{\mathcal{PT}}$. Now, unlike as in the previous construction,
the minus sign for the imaginary part is created by definition and not by
the action of $\tilde{\sigma}_{\alpha _{i}}$. In general, we may encounter
the four possibilities 
\begin{eqnarray}
\tilde{\sigma}_{\alpha _{i}}\left( \tilde{\alpha}_{j}(\varepsilon )\right)
&\in &\pm \tilde{\Delta}_{-}\qquad \text{for }\sigma _{\alpha _{i}}\left(
\alpha _{j}(\varepsilon )\right) \in \Delta _{\mp },\tilde{\alpha}%
_{j}(\varepsilon )\in \tilde{\Delta}_{+}, \\
\tilde{\sigma}_{\alpha _{i}}\left( \tilde{\alpha}_{j}(\varepsilon )\right)
&\in &\pm \tilde{\Delta}_{+}\qquad \text{for }\sigma _{\alpha _{i}}\left(
\alpha _{j}(\varepsilon )\right) \in \Delta _{\pm },\tilde{\alpha}%
_{j}(\varepsilon )\in \tilde{\Delta}_{-}.
\end{eqnarray}%
Thus any root $\tilde{\alpha}_{i}(\varepsilon )\in \tilde{\Delta}$ of the
form (\ref{def2}) is guaranteed to be mapped into $\pm \tilde{\Delta}$ under
the action of $\mathcal{W}^{\mathcal{PT}}$, which means the deformed root
system remains only invariant up to an overall sign. However, in our
application below overall signs are irrelevant so that the deformation (\ref%
{def2}) will be suitable for the application in mind.

Let us now verify that the procedure outlined above indeed leads to a closed 
$\mathcal{PT}$-symmetrically extended Weyl group $\mathcal{W}^{\mathcal{PT}}$
for some concrete examples.

\subsection{$\mathcal{PT}$-symmetric deformations of the $A_{2}$-Coxeter
group}

We recall first the action of the Weyl reflections on the simple roots by
computing (\ref{Weyl}) with the Cartan matrix $K_{ij}=2\alpha _{i}\cdot
\alpha _{j}/\alpha _{j}^{2}$, whose entries are $K_{11}=K_{22}=2$, $%
K_{12}=K_{21}=-1$. The combinations of Weyl reflections achieving a
reflection across the hyperplanes through the origin orthogonal to the three
positive roots $\alpha _{1}$, $\alpha _{2}$ and $\alpha _{1}+\alpha _{2}$ of 
$A_{2}$ are 
\begin{equation}
\begin{array}{rlll}
\sigma _{1}: & \alpha _{1}\mapsto -\alpha _{1}, & \alpha _{2}\mapsto \alpha
_{1}+\alpha _{2},~~~~ & \alpha _{1}+\alpha _{2}\mapsto \alpha _{2}, \\ 
\sigma _{2}: & \alpha _{1}\mapsto \alpha _{1}+\alpha _{2},~~~~ & \alpha
_{2}\mapsto -\alpha _{2}, & \alpha _{1}+\alpha _{2}\mapsto \alpha _{1}, \\ 
\sigma _{1}\sigma _{2}\sigma _{1}: & \alpha _{1}\mapsto -\alpha _{2}, & 
\alpha _{2}\mapsto -\alpha _{1}, & \alpha _{1}+\alpha _{2}\mapsto -\alpha
_{1}-\alpha _{2}.%
\end{array}
\label{ref}
\end{equation}
As a starting point for the deformation we choose the simple root $\alpha
_{1}$ and extend the parity symmetry across the hyperplane through the
origin orthogonal to this root. According to (\ref{hj}) the deformed root
should be taken to 
\begin{equation}
\tilde{\alpha}_{1}(\varepsilon ):=R(\varepsilon )\alpha _{1}\pm
iI(\varepsilon )\lambda _{2},  \label{a1}
\end{equation}
where we introduced the fundamental weight $\lambda _{2}=(\alpha
_{1}+2\alpha _{2})/3$. Next we compute the action of the complex reflections 
$\tilde{\sigma}_{\alpha _{i}}\mathcal{\ }$on this root in order to construct
the remaining five deformed roots. By construction we have 
\begin{equation}
\tilde{\sigma}_{1}\tilde{\alpha}_{1}(\varepsilon )=-R(\varepsilon )\alpha
_{1}\mp iI(\varepsilon )\lambda _{2}=:-\tilde{\alpha}_{1}(\varepsilon ).
\end{equation}
Having determined $\pm \tilde{\alpha}_{1}$ we may calculate the deformations
of $\alpha _{2}$ from $\mp \tilde{\sigma}_{1}\tilde{\sigma}_{2}\tilde{\sigma}%
_{1}\tilde{\alpha}_{1}$ guided by the undeformed case (\ref{ref}). We
compute 
\begin{equation}
-\tilde{\sigma}_{1}\tilde{\sigma}_{2}\tilde{\sigma}_{1}\tilde{\alpha}%
_{1}(\varepsilon )=R(\varepsilon )\alpha _{2}\mp iI(\varepsilon )\lambda
_{1}=:\tilde{\alpha}_{2}(\varepsilon )  \label{a2}
\end{equation}
where we obtained the fundamental weight $\lambda _{1}=(2\alpha _{1}+\alpha
_{2})/3$. We may verify that the remaining reflections in (\ref{ref}) indeed
yield a consistent system 
\begin{eqnarray}
\tilde{\sigma}_{2}\tilde{\alpha}_{1}(\varepsilon ) &=&R(\varepsilon )(\alpha
_{1}+\alpha _{2})\mp iI(\varepsilon )(\lambda _{1}-\lambda _{2})=\tilde{%
\alpha}_{1}(\varepsilon )+\tilde{\alpha}_{2}(\varepsilon ),  \label{d1} \\
\tilde{\sigma}_{1}\tilde{\alpha}_{2}(\varepsilon ) &=&R(\varepsilon )(\alpha
_{1}+\alpha _{2})\mp iI(\varepsilon )(\lambda _{1}-\lambda _{2})=\tilde{%
\alpha}_{1}(\varepsilon )+\tilde{\alpha}_{2}(\varepsilon ),  \label{d2} \\
\tilde{\sigma}_{2}\tilde{\alpha}_{2}(\varepsilon ) &=&-R(\varepsilon )\alpha
_{2}\pm iI(\varepsilon )\lambda _{1}=-\tilde{\alpha}_{2}(\varepsilon ),
\label{al12} \\
\tilde{\sigma}_{1}\tilde{\sigma}_{2}\tilde{\sigma}_{1}\tilde{\alpha}%
_{2}(\varepsilon ) &=&-R(\varepsilon )\alpha _{1}\mp iI(\varepsilon )\lambda
_{2}=-\tilde{\alpha}_{1}(\varepsilon ).  \label{a100}
\end{eqnarray}
We may verify that in (\ref{d1}) and (\ref{d2}) the imaginary part of the
deformed root $\tilde{\alpha}_{1}(\varepsilon )+\tilde{\alpha}%
_{2}(\varepsilon )$ is indeed orthogonal to the root $\alpha _{1}+\alpha
_{2} $ as it should be by construction. Alternatively we could have started
with the expressions (\ref{a1}) and (\ref{a2}) involving the ambiguities of
the relative signs in front of the imaginary parts and the unknown functions 
$R(\varepsilon )$ and $I(\varepsilon )$. The subsequent action of
combinations of $\tilde{\sigma}_{1}$ and $\tilde{\sigma}_{2}$ would fix the
sign ambiguity and produce the same set of deformed roots. Note that we also
have the property 
\begin{equation}
\tilde{\alpha}_{i}\cdot \tilde{\alpha}_{j}=\tilde{\sigma}_{\alpha _{k}}(%
\tilde{\alpha}_{i})\cdot \tilde{\sigma}_{\alpha _{k}}(\tilde{\alpha}%
_{j}),\qquad i,j,k=1,2.  \label{innerp}
\end{equation}
If we impose the additional constraint that inner products of root and
deformed roots are identical 
\begin{equation}
\alpha _{i}\cdot \alpha _{j}=\tilde{\alpha}_{i}\cdot \tilde{\alpha}_{j},
\label{inner}
\end{equation}
we may fix the deformation functions to $R(\varepsilon )=\cosh \varepsilon $%
, $I(\varepsilon )=\sqrt{3}\sinh \varepsilon $. The factor of $\sqrt{3}$ in
the function $I(\varepsilon )$ is somewhat natural as it ensures that the
roots in the real part of the deformed roots $\tilde{\alpha}_{i}(\varepsilon
)$ and the weights in the imaginary part have the same length. As intended,
we have achieved a simple one-to-one relation between (\ref{ref}) and the
corresponding identities for the deformed system simply by replacing $\sigma
_{i}\rightarrow \tilde{\sigma}_{i}$ and $\alpha _{i}\rightarrow \tilde{\alpha%
}_{i}(\varepsilon )$. We depict the roots and the hyperplanes in figure 1.

\begin{center}
\includegraphics[height=10.0cm]{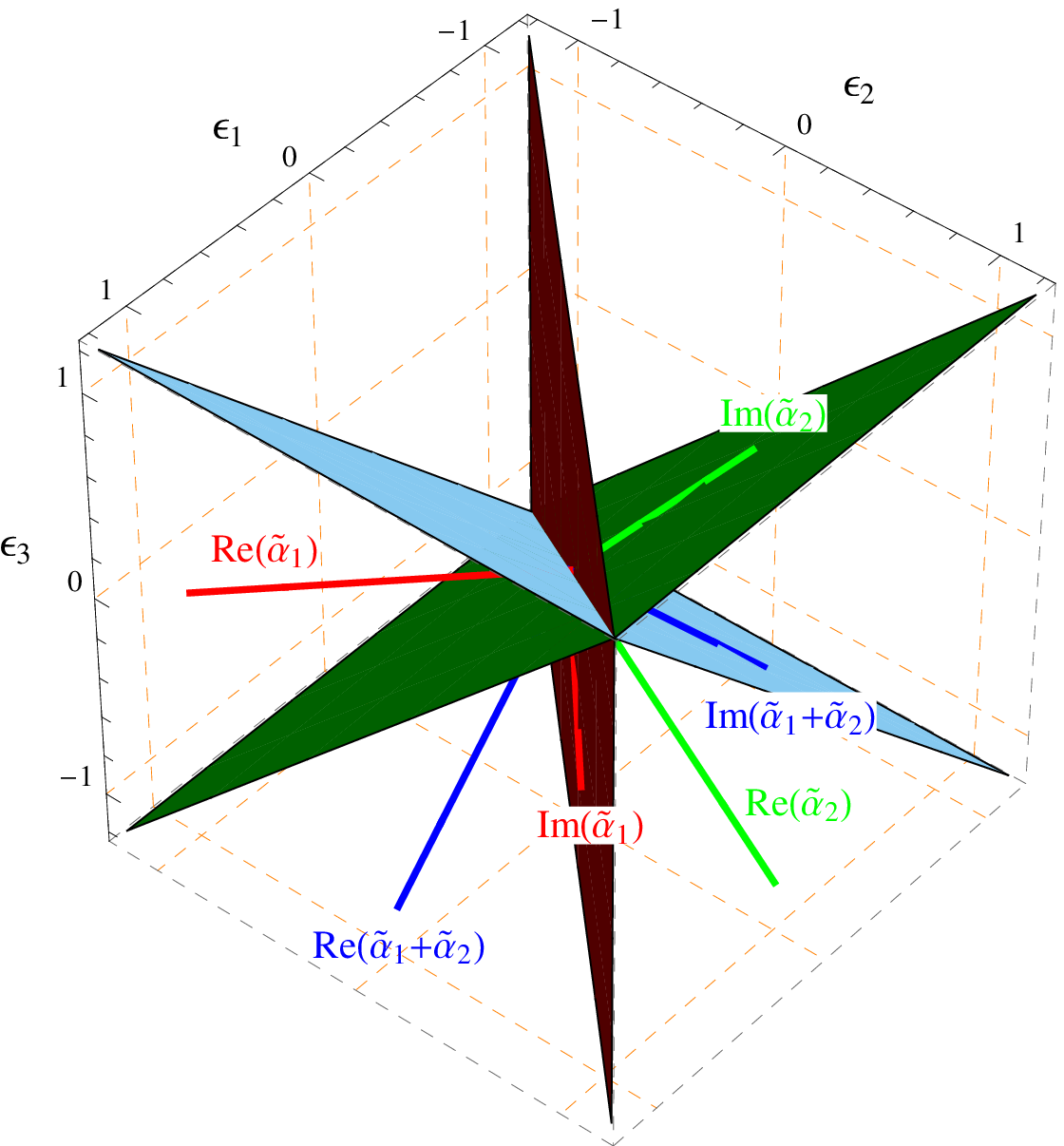}
\end{center}

\noindent {\small Figure 1: Real and imaginary parts of the $A_{2}$ deformed
roots divided by $R(\varepsilon )$ and $I(\varepsilon )$, respectively, in
the three dimensional standard representation for the simple roots $\alpha
_{1}=\varepsilon _{1}-\varepsilon _{2}$, $\alpha _{2}=\varepsilon
_{2}-\varepsilon _{3}$. Both parts of a particular positive root $\tilde{%
\alpha}_{i}$ are depicted in the same colour (on-line).}

Alternatively we can deform the six roots according to the principle (\ref%
{alt}) as 
\begin{eqnarray}
\pm \alpha _{1} &\rightarrow &\tilde{\alpha}_{1}^{\pm }=\pm R(\varepsilon
)\alpha _{1}+iI(\varepsilon )\alpha _{1} \\
\pm \alpha _{2} &\rightarrow &\tilde{\alpha}_{2}^{\pm }=\pm R(\varepsilon
)\alpha _{2}+iI(\varepsilon )\alpha _{2} \\
\pm (\alpha _{1}+\alpha _{2}) &\rightarrow &\tilde{\alpha}_{1}^{\pm }+\tilde{%
\alpha}_{2}^{\pm }=\pm R(\varepsilon )(\alpha _{1}+\alpha
_{2})+iI(\varepsilon )(\alpha _{1}+\alpha _{2}).
\end{eqnarray}%
As pointed out we no longer have $\tilde{\alpha}_{1}^{-}=-\tilde{\alpha}%
_{1}^{+}$. Nonetheless, it is easy to verify that these roots are mapped
into each other by $\mathcal{W}^{\mathcal{PT}}$ as 
\begin{equation}
\begin{array}{rlll}
\tilde{\sigma}_{1}: & \tilde{\alpha}_{1}^{+}\mapsto \tilde{\alpha}_{1}^{-},
& \tilde{\alpha}_{2}^{+}\mapsto -(\tilde{\alpha}_{1}^{-}+\tilde{\alpha}%
_{2}^{-}),\quad & \tilde{\alpha}_{1}^{+}+\tilde{\alpha}_{2}^{+}\mapsto -%
\tilde{\alpha}_{2}^{-}, \\ 
\tilde{\sigma}_{2}: & \tilde{\alpha}_{1}^{+}\mapsto -(\tilde{\alpha}_{1}^{-}+%
\tilde{\alpha}_{2}^{-}),\quad & \tilde{\alpha}_{2}^{+}\mapsto -\tilde{\alpha}%
_{2}^{-}, & \tilde{\alpha}_{1}^{+}+\tilde{\alpha}_{2}^{+}\mapsto -\tilde{%
\alpha}_{1-}^{-}, \\ 
\tilde{\sigma}_{1}\tilde{\sigma}_{2}\tilde{\sigma}_{1}: & \tilde{\alpha}%
_{1}^{+}\mapsto \tilde{\alpha}_{2}^{-}, & \tilde{\alpha}_{2}^{+}\mapsto 
\tilde{\alpha}_{1}^{-}, & \tilde{\alpha}_{1}^{+}+\tilde{\alpha}%
_{2}^{+}\mapsto \tilde{\alpha}_{1}^{-}+\tilde{\alpha}_{2}^{-}.%
\end{array}%
\end{equation}%
For these roots the inner product is not preserved and (\ref{innerp}) does
not hold in this case.

\subsection{$\mathcal{PT}$-symmetric deformations of the $G_{2}$-Coxeter
group}

Since only roots of one length emerge in root systems related to simply
laced Lie algebras, some features discussed this far are slightly different
for non-simply laced cases. Let us therefore present one explicitly example
in order to exhibit the differences. We recall, see e.g. \cite{Hum,HUm2},
that the set of roots invariant under the $G_{2}$-Coxeter group separates
into a set of short and long roots $\Delta _{s}$ and $\Delta _{l}$,
respectively, 
\begin{equation}
\Delta =\Delta _{s}\cup \Delta _{l}=\pm \{\alpha _{1},(\alpha _{1}+\alpha
_{2}),(2\alpha _{1}+\alpha _{2})\}\cup \pm \{\alpha _{2},(3\alpha
_{1}+\alpha _{2}),(3\alpha _{1}+2\alpha _{2})\}.
\end{equation}
Using the Cartan matrix with entries $K_{11}=K_{22}=2$, $K_{12}=-1$ and $%
K_{21}=-3$ we may compute the action of the Weyl reflections on the simple
roots by evaluating (\ref{Weyl}). The combinations of Weyl reflections
achieving a reflection across the hyperplanes through the origin orthogonal
to the six positive roots are presented in the following table:

\begin{center}
$ 
\begin{array}{|r||c|c|c|c|c|c|}
\hline
& \alpha _{1} & \alpha _{1}+\alpha _{2} & 2\alpha _{1}+\alpha _{2} & \alpha
_{2} & 3\alpha _{1}+\alpha _{2} & 3\alpha _{1}+2\alpha _{2} \\ \hline\hline
\sigma _{1}: & -\alpha _{1} & 2\alpha _{1}+\alpha _{2} & \alpha _{1}+\alpha
_{2} & 3\alpha _{1}+\alpha _{2} & \alpha _{2} & 3\alpha _{1}+2\alpha _{2} \\ 
\hline
\sigma _{2}: & \alpha _{1}+\alpha _{2} & \alpha _{1} & 2\alpha _{1}+\alpha
_{2} & -\alpha _{2} & 3\alpha _{1}+2\alpha _{2} & 3\alpha _{1}+\alpha _{2}
\\ \hline
\sigma _{2}\sigma _{1}\sigma _{2}: & 2\alpha _{1}+\alpha _{2} & -\alpha
_{1}-\alpha _{2} & \alpha _{1} & -3\alpha _{1}-2\alpha _{2} & 3\alpha
_{1}+\alpha _{2} & -\alpha _{2} \\ \hline
\sigma _{1}\sigma _{2}\sigma _{1}: & -2\alpha _{1}-\alpha _{2} & \alpha
_{1}+\alpha _{2} & -\alpha _{1} & 3\alpha _{1}+2\alpha _{2} & -3\alpha
_{1}-\alpha _{2} & \alpha _{2} \\ \hline
\sigma _{1}\sigma _{2}\sigma _{1}\sigma _{2}\sigma _{1}: & -\alpha
_{1}-\alpha _{2} & -\alpha _{1} & -2\alpha _{1}-\alpha _{2} & \alpha _{2} & 
-3\alpha _{1}-2\alpha _{2} & -3\alpha _{1}-\alpha _{2} \\ \hline
\sigma _{2}\sigma _{1}\sigma _{2}\sigma _{1}\sigma _{2}: & \alpha _{1} & 
-2\alpha _{1}-\alpha _{2} & -\alpha _{1}-\alpha _{2} & -3\alpha _{1}-\alpha
_{2} & -\alpha _{2} & -3\alpha _{1}-2\alpha _{2} \\ \hline
\end{array}
$
\end{center}

\noindent {\small Table 1: Simple Weyl reflections acting on the six
positive roots of $G_{2}$}

Having assembled the key properties for the undeformed root system, we
choose as a starting point for the construction of $\tilde{\Delta}$ the
deformation of the simple roots $\alpha _{1}$ or $\alpha _{2}$ and extend
the parity symmetry across the hyperplane through the origin orthogonal to
these roots. According to (\ref{hj}) the deformed counterparts can be taken
to be 
\begin{eqnarray}
\tilde{\alpha}_{1}(\varepsilon ) &=&R(\varepsilon )\alpha _{1}\pm
iI(\varepsilon )\lambda _{2}, \\
\tilde{\alpha}_{2}(\varepsilon ) &=&R(\varepsilon )\alpha _{2}\mp
i3I(\varepsilon )\lambda _{1},
\end{eqnarray}
where we used the two fundamental weights $\lambda _{1}=2\alpha _{1}+\alpha
_{2}$ and $\lambda _{2}=3\alpha _{1}+2\alpha _{2}$ of $G_{2}$. Acting now
with products of the complex reflections $\tilde{\sigma}_{\alpha _{i}}%
\mathcal{\ }$first on $\tilde{\alpha}_{1}(\varepsilon )$ yields the
deformations of the short roots 
\begin{eqnarray}
\tilde{\sigma}_{1}\tilde{\alpha}_{1}(\varepsilon ) &=&-R(\varepsilon )\alpha
_{1}\mp iI(\varepsilon )\lambda _{2}=-\tilde{\alpha}_{1}(\varepsilon ),
\label{s1} \\
\tilde{\sigma}_{2}\tilde{\alpha}_{1}(\varepsilon ) &=&R(\varepsilon )(\alpha
_{1}+\alpha _{2})\mp iI(\varepsilon )(3\lambda _{1}-\lambda _{2})=\tilde{%
\alpha}_{1}(\varepsilon )+\tilde{\alpha}_{2}(\varepsilon ), \\
\tilde{\sigma}_{1}\tilde{\sigma}_{2}\tilde{\sigma}_{1}\tilde{\alpha}%
_{1}(\varepsilon ) &=&-R(\varepsilon )(2\alpha _{1}+\alpha _{2})\mp
iI(\varepsilon )(3\lambda _{1}-2\lambda _{2})=-2\tilde{\alpha}%
_{1}(\varepsilon )-\tilde{\alpha}_{2}(\varepsilon ), \\
\tilde{\sigma}_{2}\tilde{\sigma}_{1}\tilde{\sigma}_{2}\tilde{\alpha}%
_{1}(\varepsilon ) &=&R(\varepsilon )(2\alpha _{1}+\alpha _{2})\mp
iI(\varepsilon )(3\lambda _{1}-2\lambda _{2})=2\tilde{\alpha}%
_{1}(\varepsilon )+\tilde{\alpha}_{2}(\varepsilon ), \\
\tilde{\sigma}_{1}\tilde{\sigma}_{2}\tilde{\sigma}_{1}\tilde{\sigma}_{2}%
\tilde{\sigma}_{1}\tilde{\alpha}_{1}(\varepsilon ) &=&-R(\varepsilon
)(\alpha _{1}+\alpha _{2})\pm iI(\varepsilon )(3\lambda _{1}-\lambda _{2})=-%
\tilde{\alpha}_{1}(\varepsilon )-\tilde{\alpha}_{2}(\varepsilon ), \\
\tilde{\sigma}_{2}\tilde{\sigma}_{1}\tilde{\sigma}_{2}\tilde{\sigma}_{1}%
\tilde{\sigma}_{2}\tilde{\alpha}_{1}(\varepsilon ) &=&R(\varepsilon )\alpha
_{1}\pm iI(\varepsilon )\lambda _{2}=\tilde{\alpha}_{1}(\varepsilon ).
\label{s6}
\end{eqnarray}
The action of products of reflections $\tilde{\sigma}_{\alpha _{i}}\mathcal{%
\ }$on $\tilde{\alpha}_{2}(\varepsilon )$ yields the deformations of the
long roots 
\begin{eqnarray}
\tilde{\sigma}_{1}\tilde{\alpha}_{2}(\varepsilon ) &=&R(\varepsilon
)(3\alpha _{1}+\alpha _{2})\mp i3I(\varepsilon )(\lambda _{1}-\lambda _{2})=3%
\tilde{\alpha}_{1}(\varepsilon )+\tilde{\alpha}_{2}(\varepsilon ),
\label{l1} \\
\tilde{\sigma}_{2}\tilde{\alpha}_{2}(\varepsilon ) &=&-R(\varepsilon )\alpha
_{2}\pm i3I(\varepsilon )\lambda _{1}=-\tilde{\alpha}_{2}(\varepsilon ), \\
\tilde{\sigma}_{1}\tilde{\sigma}_{2}\tilde{\sigma}_{1}\tilde{\alpha}%
_{2}(\varepsilon ) &=&R(\varepsilon )(3\alpha _{1}+2\alpha _{2})\mp
i3I(\varepsilon )(2\lambda _{1}-\lambda _{2})=3\tilde{\alpha}%
_{1}(\varepsilon )+2\tilde{\alpha}_{2}(\varepsilon ), \\
\tilde{\sigma}_{2}\tilde{\sigma}_{1}\tilde{\sigma}_{2}\tilde{\alpha}%
_{2}(\varepsilon ) &=&-R(\varepsilon )(3\alpha _{1}+2\alpha _{2})\pm
i3I(\varepsilon )(2\lambda _{1}-\lambda _{2})=-3\tilde{\alpha}%
_{1}(\varepsilon )-2\tilde{\alpha}_{2}(\varepsilon ),~~~ \\
\tilde{\sigma}_{1}\tilde{\sigma}_{2}\tilde{\sigma}_{1}\tilde{\sigma}_{2}%
\tilde{\sigma}_{1}\tilde{\alpha}_{2}(\varepsilon ) &=&R(\varepsilon )\alpha
_{2}\mp i3I(\varepsilon )\lambda _{1}=\tilde{\alpha}_{2}(\varepsilon ), \\
\tilde{\sigma}_{2}\tilde{\sigma}_{1}\tilde{\sigma}_{2}\tilde{\sigma}_{1}%
\tilde{\sigma}_{2}\tilde{\alpha}_{2}(\varepsilon ) &=&-R(\varepsilon
)(3\alpha _{1}+\alpha _{2})\pm i3I(\varepsilon )(\lambda _{1}-\lambda
_{2})=-3\tilde{\alpha}_{1}(\varepsilon )-\tilde{\alpha}_{2}(\varepsilon ).
\label{l6}
\end{eqnarray}
For a particular representation we depict the constructed roots in figure 2.

\noindent \includegraphics[width=8.0cm,height=8.0cm]{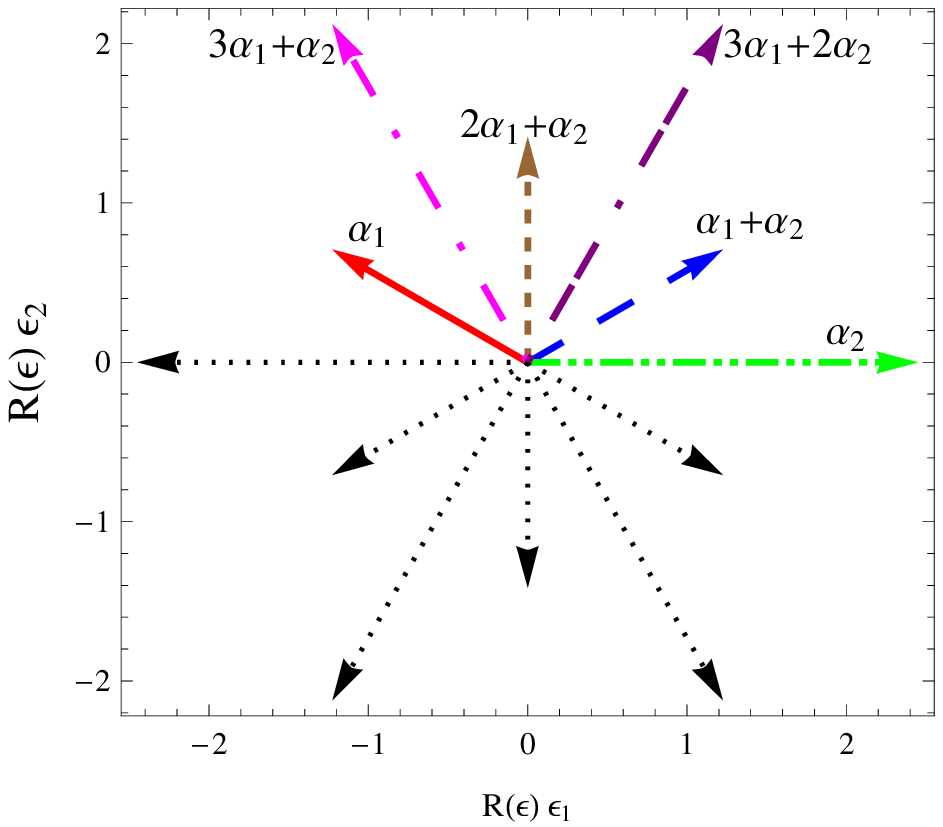} %
\includegraphics[width=8.0cm,height=8.0cm]{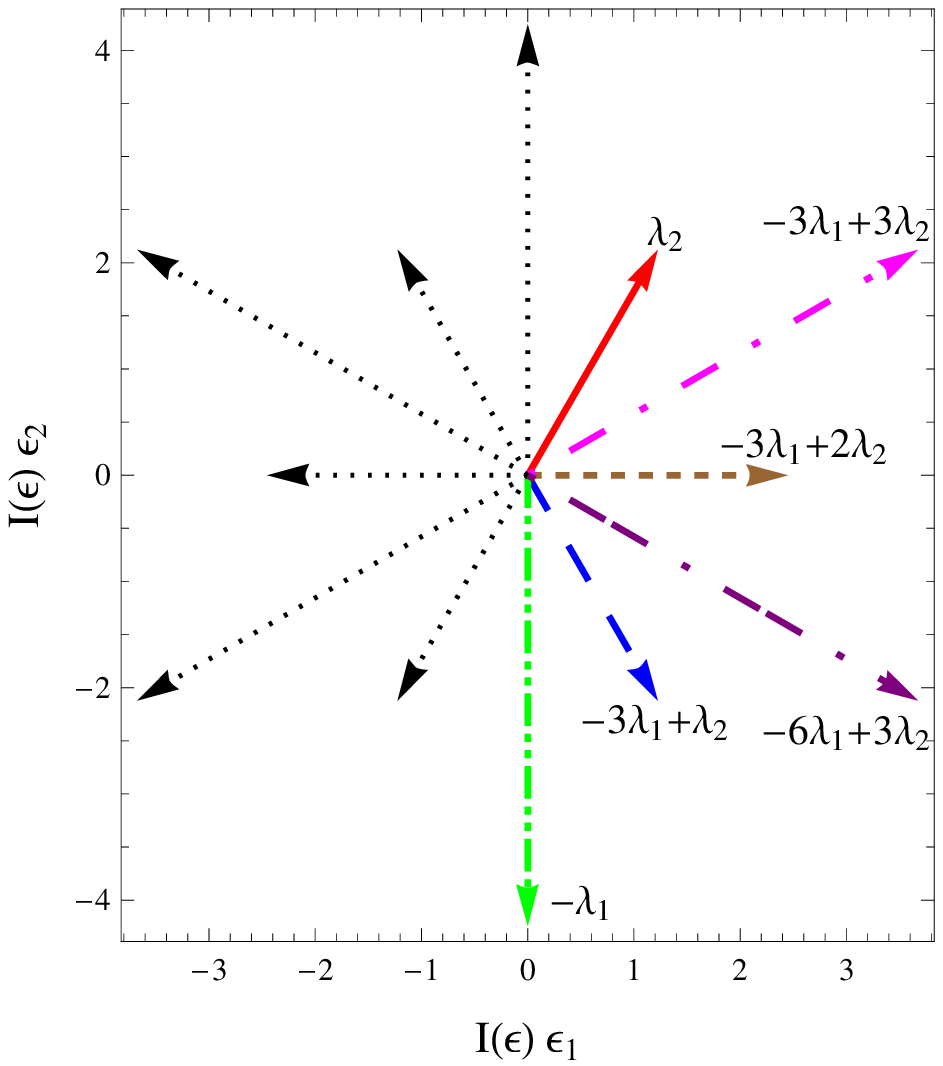}

\smallskip

\noindent {\small Figure 2: Real and imaginary parts of the $G_{2}$-deformed
roots in the two dimensional basis for the simple roots $\alpha
_{1}=(\varepsilon _{2}-\sqrt{3}\varepsilon _{1})/\sqrt{2}$, $\alpha _{2}=%
\sqrt{6}\varepsilon _{1}$. Both parts of a particular positive root $\tilde{%
\alpha}_{i}$ are depicted in the same line style (on-line also colour).}

As it should be by construction, we can check for consistence once more that
indeed the imaginary part is orthogonal to the real part of each deformed
root. Again we observe the property 
\begin{equation}
\tilde{\alpha}_{i}\cdot \tilde{\alpha}_{j}=\tilde{\sigma}_{\alpha _{k}}(%
\tilde{\alpha}_{i})\cdot \tilde{\sigma}_{\alpha _{k}}(\tilde{\alpha}%
_{j}),\qquad i,j,k=1,2
\end{equation}
and with the additional requirement 
\begin{equation}
\alpha _{i}\cdot \alpha _{j}=\tilde{\alpha}_{i}\cdot \tilde{\alpha}_{j},
\end{equation}
we may fix the deformation functions to $R(\varepsilon )=\cosh \varepsilon $%
, $I(\varepsilon )=1/\sqrt{3}\sinh \varepsilon $. We have achieved a simple
one-to-one relation between (\ref{ref}) and the corresponding identities for
the deformed system simply by replacing $\sigma _{i}\rightarrow \tilde{\sigma%
}_{i}$ and $\alpha _{i}\rightarrow \tilde{\alpha}_{i}(\varepsilon )$.

Clearly we can also choose the deformation according to (\ref{def2}) as for
the $A_{2}$-case, but we will not report this here.

\section{$\mathcal{PT}$-symmetric deformations of Calogero-Moser-Sutherland
models}

Taking the previous remarks into account it is now straightforward to
formulate new types of CMS-models, which are invariant under the action of $%
\mathcal{PT}$-symmetrically extended Weyl groups $\mathcal{W}^{\mathcal{PT}}$%
. The Hamiltonian will be of the form $\mathcal{H}_{\mathcal{PT}\text{CMS}}$
as specified in (\ref{PTCM}). Let us study some concrete examples.

\subsection{$\mathcal{PT}$-symmetrically deformed $A_{2}$%
-Calogero-Moser-Sutherland models}

Beyond the two particle problem the $A_{2}$-CMS model is the simplest
classical example, constituting in some representation the three-body
problem with a two particle interaction \cite{Cal3}. For this Coxeter group
we consider now the Hamiltonian $\mathcal{H}_{\mathcal{PT}\text{CMS}}$ in (%
\ref{PTCM}) with the two simple roots taken in the standard three
dimensional representation $\alpha _{1}=\varepsilon _{1}-\varepsilon _{2}$
and $\alpha _{2}=\varepsilon _{2}-\varepsilon _{3}$, with $\varepsilon _{i}$
being an orthogonal basis in $\mathbb{R}^{3}$ with $\varepsilon _{i}\cdot
\varepsilon _{j}=\delta _{ij}$ and the dynamical variables to be $%
q=\{q_{1},q_{2},q_{3}\}$. Using then the deformed roots as constructed in (%
\ref{a1})-(\ref{a100}), the potential of the $\mathcal{PT}$-symmetrically
extended model acquires the form 
\begin{equation}
V_{\mathcal{PT}\text{CMS}}^{A_{2}}=g\dsum\limits_{\substack{ 1\leq j<k\leq 3 
\\ j,k\neq l}}V[R(\varepsilon )(q_{j}-q_{k})+i(-1)^{j+k}I(\varepsilon
)(q_{j}+q_{k}-2q_{l})],  \label{A21}
\end{equation}%
where $V(x)$ can be of Calogero type, i.e. $V(x)=1/x^{2}$ or any of the
functions $1/\sin ^{2}x$, $1/\sinh ^{2}x$, $1/\func{sn}^{2}x$.

By construction these potentials are symmetric with regard to $\mathcal{W}^{%
\mathcal{PT}}$, which of course can also be seen explicitly for the
dynamical variables $\tilde{\sigma}_{\alpha _{1}}\equiv $ $%
q_{1}\leftrightarrow q_{2},i\rightarrow -i$, $\tilde{\sigma}_{\alpha
_{2}}\equiv $ $q_{2}\leftrightarrow q_{3},i\rightarrow -i$ and $\tilde{\sigma%
}_{\alpha _{1}+\alpha _{2}}\equiv $ $q_{1}\leftrightarrow q_{3},i\rightarrow
-i$.

Instead of the three dimensional representation we may also represent the
roots in a two dimensional space, i.e. $\alpha _{1}=\sqrt{2}\varepsilon _{1}$%
, $\alpha _{2}=\sqrt{3/2}\varepsilon _{2}-\sqrt{2}\varepsilon _{1}$ and
express the dynamical variables in terms of Jacobi relative coordinates $%
q=\{X,Y\}$. Comparison between the two representations then leads to the
well known relations between the different sets of variables $%
X=(q_{1}-q_{2})/\sqrt{2}$ and $Y=(q_{1}+q_{2}-2q_{3})/\sqrt{6}$. The third
coordinate is usually taken to be the center-of-mass coordinate $%
R=(q_{1}+q_{2}+q_{3})/3$. Moreover, it is convenient to parameterize$~X$ and 
$Y$ further in terms of polar coordinates $X$ $=r\sin \phi $, $Y=r\cos \phi $%
. In this formulation the relations for the potential simplify even more
with the special choice $R(\varepsilon )=\cosh \varepsilon $ and $%
I(\varepsilon )=\sqrt{3}\sinh \varepsilon $ as already mentioned after (\ref%
{inner}). With these choices the potential (\ref{A21}) is transformed into 
\begin{equation}
V_{\mathcal{PT}\text{CMS}}^{A_{2}}=g\dsum\limits_{k=-1,0,1}V\left[ \sqrt{2}%
r\sin (\phi -i\varepsilon +k\frac{2\pi }{3})\right] .  \label{MI}
\end{equation}
Taking the special case $V(x)=g/x^{2}$ the version (\ref{MI}) of the $%
\mathcal{PT}$-symmetrically extended $A_{2}$-Calogero model is essentially
the potential suggested in \cite{Milos}, where it was obtained by deforming
directly the Calogero model in the form (\ref{MI}) for $\varepsilon =0$
across the symmetry $\phi \rightarrow -\phi $ via the recipe $\phi \mapsto
\phi -i\varepsilon $. We have demonstrated here how to obtain it as a
special case from a more general and systematic setting. The virtue of the
version (\ref{MI}) in the new coordinate system is that it leads to a
separable Schr\"{o}dinger equation. In section 4.3 we make use of this fact
and investigate some properties of the model, notably to construct its
eigenfunctions and eigenvalues.

Clearly we may also choose the deformations according to the alternative
deformation (\ref{def2}), in which case the $\mathcal{PT}$-symmetrically
extended model is of the form 
\begin{equation}
V_{\mathcal{PT}\text{CMS}}^{A_{2}}=\frac{g}{2}\dsum\limits_{1\leq j<k\leq
3}V[(R(\varepsilon )+iI(\varepsilon ))(q_{j}-q_{k})]+\frac{g}{2}%
\dsum\limits_{1\leq j<k\leq 3}V[(R(\varepsilon )-iI(\varepsilon
))(q_{j}-q_{k})],  \label{zz}
\end{equation}
when choosing the roots to be in the standard representation. Note that,
whereas in the undeformed case the contributions form any negative roots
equals the one resulting from its positive counterpart, now these roots give
different contributions. Expressing (\ref{zz}) in terms of Jacobian relative
coordinates and making in addition the choice $R(\varepsilon )=1$ and $%
I(\varepsilon )=\varepsilon /r$ the potential simply becomes 
\begin{equation}
V_{\mathcal{PT}\text{CMS}}^{A_{2}}=\frac{g}{2}\dsum\limits_{k=0,\pm 1}\left[
V\left[ \sqrt{2(}r+i\varepsilon )\sin (\phi +\frac{2\pi }{3}k)\right] +V%
\left[ \sqrt{2(}r-i\varepsilon )\sin (\phi +\frac{2\pi }{3}k)\right] \right]
.  \label{rr}
\end{equation}
Note that in the choice for $I(\varepsilon )$ we made use fact that we can
multiply this quantity by any invariant of $\mathcal{W}^{\mathcal{PT}}$.
Clearly $r=\sqrt{(\alpha _{1}\cdot q)^{2}/3+(\alpha _{2}\cdot
q)^{2}/3+(\alpha _{1}\cdot q+\alpha _{2}\cdot q)^{2}/3}$ is such an
invariant. Thus when we restrict the sum in (\ref{zz}), (\ref{rr}) to the
positive or negative roots only the deformation is simply achieved by $%
r\mapsto r+i\varepsilon $ or $r\mapsto r-i\varepsilon $, respectively. This
corresponds to the deformation of the symmetry $r\rightarrow -r$. One should
note that the restriction to just half of the number of roots will break the
invariance under the action of $\mathcal{W}^{\mathcal{PT}}$.

\subsection{$\mathcal{PT}$-symmetrically deformed $G_{2}$%
-Calogero-Moser-Sutherland models}

The $G_{2}$-CMS-model, constitutes a further standard example, since it can
be viewed as the classical three-body problem with a two and a three-body
interaction term \cite{Wolf}. As in the previous subsection we may now
realize the roots in various different ways. Either we can take the
so-called standard three dimensional representation for the simple roots $%
\alpha _{1}=\varepsilon _{1}-\varepsilon _{2}$, $\alpha _{2}=-2\varepsilon
_{1}+\varepsilon _{2}+\varepsilon _{3}$ as concrete realization for the
simple roots of $G_{2}$ in $\mathbb{R}^{3}$ and the dynamical variables to
be $q=\{q_{1},q_{2},q_{3}\}$ or alternatively we may also represent them in
a two dimensional space as $\alpha _{1}=(\varepsilon _{2}-\sqrt{3}%
\varepsilon _{1})/\sqrt{2}$, $\alpha _{2}=\sqrt{6}\varepsilon _{1}$ and
express the dynamical variables in terms of Jacobi relative coordinates $%
q=\{X,Y\}$. \ Once again the comparison between the two representations
yields to the same relations for the Jacobi relative coordinates $\
X=(q_{1}-q_{2})/\sqrt{2}$ and $Y=(q_{1}+q_{2}-2q_{3})/\sqrt{6}$. Explicitly
the inner products in all coordinate systems are computed to 
\begin{eqnarray}
\alpha _{1}\cdot q &=&q_{1}-q_{2}=\sqrt{2}X=\sqrt{2}r\sin \phi ,  \label{t1}
\\
(\alpha _{1}+\alpha _{2})\cdot q &=&q_{3}-q_{1}=-\frac{1}{\sqrt{2}}(\sqrt{3}%
Y+X)=-\sqrt{2}r\sin (\frac{2\pi }{3}-\phi ),  \label{t2} \\
(2\alpha _{1}+\alpha _{2})\cdot q &=&q_{3}-q_{2}=-\frac{1}{\sqrt{2}}(\sqrt{3}%
Y-X)=-\sqrt{2}r\sin (\frac{2\pi }{3}+\phi ),  \label{t3} \\
\alpha _{2}\cdot q &=&q_{2}+q_{3}-2q_{1}=-\sqrt{\frac{3}{2}(}\sqrt{3}X+Y)=%
\sqrt{6}r\cos (\frac{2\pi }{3}+\phi ), \\
(3\alpha _{1}+\alpha _{2})\cdot q &=&q_{1}+q_{3}-2q_{2}=\sqrt{\frac{3}{2}(}%
\sqrt{3}X-Y)=\sqrt{6}r\cos (\frac{2\pi }{3}-\phi ), \\
(3\alpha _{1}+2\alpha _{2})\cdot q &=&2q_{3}-q_{1}-q_{2}=-\sqrt{6}Y=-\sqrt{6}%
r\cos \phi .  \label{t6}
\end{eqnarray}
The expressions for the short roots (\ref{t1}), (\ref{t2}) and (\ref{t3})
just yield the expressions for the $A_{2}$-roots $\alpha _{1}$, $-\alpha
_{2} $ and $-\alpha _{1}-\alpha _{2}$ in the standard representation. Using
the expressions (\ref{t1})-(\ref{t6}) in the Hamiltonian $\mathcal{H}_{%
\mathcal{PT}\text{CMS}}$ in (\ref{PTCM}), the $\mathcal{PT}$-symmetrically
deformed $G_{2}$-CMS potential becomes 
\begin{eqnarray}
V_{\mathcal{PT}\text{CMS}}^{G_{2}} &=&g_{s}\dsum\limits_{\substack{ 1\leq
j<k\leq 3  \\ j,k\neq l}}V[R(\varepsilon
)(q_{j}-q_{k})+i/3(-1)^{j+k}I(\varepsilon )(q_{j}+q_{k}-2q_{l})] \\
&&+g_{l}\dsum\limits_{\substack{ 1\leq j<k\leq 3  \\ j,k\neq l}}%
V[(-1)^{j+k+1}R(\varepsilon )(q_{j}+q_{k}-2q_{l})+iI(\varepsilon
)(q_{j}-q_{k})].  \notag
\end{eqnarray}
As a result of the aforementioned relation between the $A_{2}$ and $G_{2}$%
-roots the corresponding potentials reduce as $V_{\mathcal{PT}\text{CMS}%
}^{G_{2}}\rightarrow V_{\mathcal{PT}\text{CMS}}^{A_{2}}$, when we switch off
the three particle interaction $g_{l}\rightarrow 0$ and scale the
deformation function. When specifying further $R(\varepsilon )=\cosh
\varepsilon $ and $I(\varepsilon )=\sqrt{3}\sinh \varepsilon $ we obtain 
\begin{equation}
V_{\mathcal{PT}\text{CMS}}^{G_{2}}=\dsum\limits_{k=-1,0,1}g_{s}V\left[ \sqrt{%
2}r\sin (\phi -i\varepsilon +k\frac{2\pi }{3})\right] +g_{l}V\left[ \sqrt{6}%
r\cos (\phi -i\varepsilon +k\frac{2\pi }{3})\right] .
\end{equation}

Once again we may also choose a different type of deformations according to (%
\ref{def2}), in which the $\mathcal{PT}$-symmetrically extended model can be
brought into the form 
\begin{eqnarray}
V_{\mathcal{PT}\text{CMS}}^{G_{2}} &=&\frac{g_{s}}{2}\dsum\limits_{1\leq
j<k\leq 3}\left[ V[(R(\varepsilon )+iI(\varepsilon
))(q_{j}-q_{k})]+V[(R(\varepsilon )-iI(\varepsilon ))(q_{j}-q_{k})]\right] \\
&&\!\!\!\!\!\!\!\!\!\!\!\!\!\!\!\!\!\!+\frac{g_{l}}{2}\dsum\limits 
_{\substack{ 1\leq j<k\leq 3  \\ j,k\neq l}}\left[ V[(R(\varepsilon
)+iI(\varepsilon ))(q_{j}+q_{k}-2q_{l})]+V[(R(\varepsilon )+iI(\varepsilon
))(q_{j}+q_{k}-2q_{l})]\right]  \notag
\end{eqnarray}
when choosing the roots to be in the standard representation. We may also
express this in terms of Jacobian relative coordinates with the choice $%
R(\varepsilon )=1$ and $I(\varepsilon )=\varepsilon /r$ as in the $A_{2}$%
-case, such that the potential becomes 
\begin{equation}
V_{\mathcal{PT}\text{CMS}}^{G_{2}}=\dsum\limits_{\substack{ k=-1,0,1  \\ %
n=-1,1}}\frac{g_{s}}{2}V\left[ \sqrt{2}(r+i\varepsilon n)\sin (\phi +k\frac{%
2\pi }{3})\right] +\frac{g_{l}}{2}V\left[ \sqrt{6}(r+i\varepsilon n)\cos
(\phi +k\frac{2\pi }{3})\right] .  \label{alte}
\end{equation}

Let us now study some physical properties of these models.

\subsection{Eigensystems}

Let us now specialize the potential to the one of the Calogero model, i.e.
we take it to be $V(x)\sim 1/x^{2}$, and determine the eigensystems of the
deformed models. In general this is a difficult task as even for the
undeformed CMS-models the eigenfunctions are combinations of Vandermode
determinants and Jack polynomials, e.g. \cite{baker}. However, in the cases
under consideration we can follow a different route and be very explicit for
some very particular choices of the deformation functions. As illustrated in
the last subsection we may just consider the $G_{2}$-Calogero model and
treat the $A_{2}$-Calogero model as a special case of the former by
switching off the three particle interaction. The $A_{2}$-model was already
solved by Calogero \cite{Cal3} almost fourty years ago and the $G_{2}$-case
thereafter by Wolfes \cite{Wolf}. Relying on these solutions, the
construction of eigensystems for some specific deformed system is fairly
simple, as they may be obtained by implementing a shift as was done in the $%
A_{2}$-case \cite{Milos}. For other choices of the functions $R(\varepsilon
) $ and $I(\varepsilon )$ the solutions can not be constructed in direct
analogy to the undeformed case.

However, as was observed in \cite{Znojil:1999qt,Milos} even the simpler
scenario is instructive as there are a few differences in the argumentation
leading to various constraints on the parameters resulting from the
implementation of physical requirements. The main consequence of the
deformation is that some irregular solutions, which had to be discarded in
the undeformed case become perfectly viable regularized solutions after the
deformation. As a result the energy spectra of the deformed systems differ
from those of the undeformed ones. Let us briefly recall the argumentation
of \cite{Cal3,Wolf} and treat thereafter the deformed case.

\subsubsection{The undeformed case}

The above mentioned variable transformations $(x_{1},x_{2},x_{3})\rightarrow
(R,X,Y)\rightarrow (R,r,\phi )$ have the virtue that they convert the
differential equation into a form allowing for completely separability \cite%
{Cal3,Wolf}. The Laplace operator transforms simply as 
\begin{equation}
\Delta _{x_{1}x_{2}x_{3}}\rightarrow \frac{1}{3}\frac{\partial ^{2}}{%
\partial R^{2}}+\frac{\partial ^{2}}{\partial X^{2}}+\frac{\partial ^{2}}{%
\partial Y^{2}}\rightarrow \frac{1}{3}\frac{\partial ^{2}}{\partial R^{2}}+%
\frac{\partial ^{2}}{\partial r^{2}}+\frac{1}{r^{2}}\frac{\partial ^{2}}{%
\partial \phi ^{2}}+\frac{1}{r}\frac{\partial }{\partial r}
\end{equation}
the confining potential transforms as 
\begin{equation}
\frac{m^{2}}{16}\sum_{\alpha \in \Delta }(\alpha \cdot q)^{2}\rightarrow 
\frac{3}{8}m^{2}(X^{2}+Y^{2})\rightarrow \frac{\omega ^{2}}{2}r^{2}
\end{equation}
and the Calogero potential as 
\begin{eqnarray}
\frac{g_{s}}{2}\sum\limits_{\alpha \in \Delta _{s}}\frac{1}{(\alpha \cdot
q)^{2}} &\rightarrow &\frac{9}{2}g\frac{(X^{2}+Y^{2})^{2}}{%
(X^{3}-3XY^{2})^{2}}\rightarrow \frac{9}{2}\frac{g}{r^{2}\sin (3\phi )}, \\
\frac{g_{l}}{2}\sum\limits_{\alpha \in \Delta _{s}}\frac{1}{(\alpha \cdot
q)^{2}} &\rightarrow &\frac{9}{2}g\frac{(X^{2}+Y^{2})^{2}}{%
(Y^{3}-3YX^{2})^{2}}\rightarrow \frac{9}{2}\frac{g}{r^{2}\cos (3\phi )}.
\end{eqnarray}
Assembling these expressions into a Hamiltonian it is then easy to see that
in the $(R,r,\phi )$-system the eigenfunctions can be factorized into $\Psi
(R,r,\phi )=\Phi (R)\chi (r)f(\phi )$, which leads, after separating off the
center of mass motion, to the two separate eigenvalue equations 
\begin{eqnarray}
\left( -\frac{\partial ^{2}}{\partial r^{2}}-\frac{1}{r}\frac{\partial }{%
\partial r}+\omega ^{2}r^{2}+\frac{\lambda ^{2}}{r^{2}}\right) \chi (r)
&=&E\chi (r),  \label{se1} \\
\left( -\frac{\partial ^{2}}{\partial \phi ^{2}}+\frac{9g_{s}}{\sin (3\phi )}%
+\frac{9g_{l}}{\cos (3\phi )}\right) f(\phi ) &=&\lambda ^{2}f(\phi ).
\label{se2}
\end{eqnarray}
These equations may be solved generically for any real values of the
parameters $r,\phi ,g_{s},g_{l},\omega $ including even the eigenvalues $E$
and $\lambda ^{2}$ by 
\begin{eqnarray}
\chi (r) &=&r^{\lambda }\exp \left( -\omega r^{2}/2\right) \;_{1}F_{1}\left[ 
\frac{1}{2}(1+\lambda )-\frac{E}{4\omega };1+\lambda ;\omega r^{2}\right] ,
\label{r} \\
f(\phi ) &=&\sin ^{2\kappa _{s}}(3\phi )\cos ^{2\kappa _{l}}(3\phi
)\;_{2}F_{1}\left[ \kappa _{s}+\kappa _{l}-\frac{\lambda }{6},\kappa
_{s}+\kappa _{l}+\frac{\lambda }{6};2\kappa _{s}+\frac{1}{2};\sin ^{2}(3\phi
)\right] .\qquad  \label{phi}
\end{eqnarray}
Here we abbreviated the constants $\kappa _{s/l}=\kappa _{s/l}^{\pm }=(1\pm 
\sqrt{1+4g_{s/l}})/4$, $_{1}F_{1}$ denotes the Kummer confluent
hypergeometric function and $_{2}F_{1}$ the Gauss hypergeometric function.
Implementing now various different physical requirements leads to the
quantization condition for the eigenvalues and several restrictions on the
parameters 
\begin{eqnarray}
P1 &:&\qquad E=2\left\vert \omega \right\vert (2n+\lambda +1)\qquad \text{%
for }n\in \mathbb{N}_{0}, \\
P2 &:&\qquad \lambda >0, \\
P3 &:&\qquad \kappa _{s}\rightarrow \kappa _{s}^{+},\kappa _{l}\rightarrow
\kappa _{l}^{+}, \\
P4 &:&\qquad \lambda =6(\kappa _{s}+\kappa _{l}+\ell )\qquad \ \ \ \ \ \text{%
for }\ell \in \mathbb{N}_{0}.
\end{eqnarray}
We briefly recall and extend the argumentations in order to illustrate how
they need to be modified in the deformed scenario.

\paragraph{$P1:$}

The quantization condition $P1$ originates from the physical requirement
that the wavefunction should vanish for $r\rightarrow \infty $. Using the
asymptotic expansion for Kummer's confluent hypergeometric function, see
e.g. \cite{MIT}, 
\begin{eqnarray}
\;_{1}F_{1}\left[ \alpha ;\gamma ;z\right] &\sim &\frac{\Gamma (\gamma )}{%
\Gamma (\alpha )}e^{z}z^{\alpha -\gamma }G(1-\alpha ;\gamma -\alpha ,z)\text{%
\qquad\ \ \ \ \ \ \ \ \ for }\func{Re}z>0,  \label{g1} \\
_{1}F_{1}\left[ \alpha ;\gamma ;z\right] &\sim &\frac{\Gamma (\gamma )}{%
\Gamma (\gamma -\alpha )}(-z)^{-\alpha }G(\alpha ;\alpha -\gamma -1,-z)\text{%
\qquad for }\func{Re}z<0,  \label{g2}
\end{eqnarray}
with $G(\alpha ;\gamma ,z)=1+\alpha /z+\alpha (\alpha +1)\gamma (\gamma
+1)/2!/z^{2}+\ldots $, one observes that for the arguments of the solution $%
\chi (r)$ in (\ref{r}) the function will usually diverge exponentially,
unless this divergence is compensated by a diverging gamma function, either
from the corresponding $\Gamma (\alpha )$ in (\ref{g1}) or $\Gamma (\gamma
-\alpha )$ in (\ref{g2}). As this is the case when the first argument in $%
_{1}F_{1}$ becomes a negative integer, i.e. when the hypergeometric series
terminates, the wavefunction $\chi (r)$ vanishes at infinity with the
condition $P1$. For these values the Kummer confluent hypergeometric
function reduces to a generalized Laguerre polynomial $L_{n}^{\alpha }(z)$
by means of the identity 
\begin{equation}
_{1}F_{1}\left[ -n;\alpha +1;z\right] =\frac{\Gamma (n+1)\Gamma (\alpha +1)}{%
\Gamma (n+\alpha +1)}L_{n}^{\alpha }(z)\qquad \text{for }n\in \mathbb{N}_{0}%
\text{, }\alpha \in \mathbb{R}
\end{equation}
and one obtains, up to normalization, the expression for $\chi (r)$ already
found by Calogero \cite{Cal3}. Note that this argumentation does not change
even if we continue $r$ into the complex plane and $P1$ remains also valid
in that case.

\paragraph{$P2:$}

The constraint $P2$ arises from the condition that a proper physical
wavefunction should be finite on its domain. In the undeformed case the
divergence of $\chi (r)$ at $r=0$ can be cured by the constraint $P2$.
Clearly this constraint can be removed if $r$ acquires a nonvanishing
imaginary part, since the factor $r^{\lambda }$ no longer diverges for $%
r\rightarrow 0$.

\paragraph{$P3:$}

The constraint $P3$ results from same requirement as $P2$, but demanding
finiteness in the entire domain also for its derivative. For $\kappa
_{s}=\kappa _{s}^{-}$ and $\kappa _{l}=\kappa _{l}^{-}$ the prefactors in (%
\ref{phi}) would diverge for $\phi =0,\pi /3,\ldots $ and $\phi =\pi /6,\pi
/2,\ldots $, respectively. Clearly when $\func{Im}\phi \neq 0$ there is no
longer any justification for this constraint and it can be removed, thus
allowing the values $\lambda <0$.

\paragraph{$P4:$}

The quantization condition $P4$ stems from the divergence of $f(\phi )$ at
for instance $\phi =\pi /6$. This is seen from the fact that for generic
arguments the function $_{2}F_{1}\left[ \alpha ,\beta ;\gamma ;1\right] $ is
absolutely convergent when $\func{Re}\gamma >\func{Re}(\alpha +\beta )$,
which for the values in (\ref{phi}) translates into $\kappa _{l}<1/4$.
Having already excluded $\kappa _{l}^{-}$ by condition $P3$ this inequality
can never be satisfied. However, when $\alpha $ becomes a negative integer
the hypergeometric series terminates and reduces to a Jacobi polynomial $%
P_{\ell }^{\alpha ,\beta }(z)$ by means of the identity 
\begin{equation}
_{2}F_{1}\left[ -\ell ,\alpha +\beta +\ell +1;\alpha +1;z\right] =\frac{%
\Gamma (\ell +1)\Gamma (\alpha +1)}{\Gamma (\ell +\alpha +1)}P_{\ell
}^{\alpha ,\beta }(1-2z)\qquad \text{for }\ell \in \mathbb{N}_{0}\text{, }%
\alpha ,\beta \in \mathbb{R}.  \label{23}
\end{equation}
Since $P_{\ell }^{\alpha ,\beta }(-1)=(\beta +1)_{\ell }/\ell !$ with $%
(x)_{n}:=x(x+1)(x+2)\ldots (x+n)$ the divergence is removed by condition $P4$%
. Alternatively we could also equate the second argument in (\ref{phi}) to
an integer and deduce $\lambda =-6(\kappa _{s}+\kappa _{l}+\ell )$, which is
however excluded by condition $P2$. Notice that when $\func{Im}\phi \neq 0$,
we will even leave the unit circle $|z|\leq 1$, in which convergence can be
achieved unless we restrict the real part of $\phi $ depending on its
imaginary part, which seems very artificial. Thus in this case terminating
the series by means of property (\ref{23}) appears even more natural than in
the undeformed case.

In summary, when the physical constraints $P1,P2,P3,P4$ hold, the
corresponding wave functions are 
\begin{eqnarray}
\chi _{n}^{\lambda }(r) &\sim &\Gamma (n+1)\omega ^{\frac{\lambda }{2}%
}r^{\lambda }\exp \left( -\omega r^{2}/2\right) \;L_{n}^{\lambda }\left(
\omega r^{2}\right) ,  \label{phi1} \\
f_{\ell }^{\kappa _{s},,\kappa _{l}}(\phi ) &\sim &\Gamma (\ell +1)\sin
^{2\kappa _{s}}(3\phi )\cos ^{2\kappa _{l}}(3\phi )\;P_{\ell }^{2\kappa
_{s}-1/2,2\kappa _{l}-1/2}[1-2\sin ^{2}(3\phi )].\qquad  \label{phi2}
\end{eqnarray}
with energy spectrum 
\begin{equation}
E_{n\ell }=2|\omega |\left[ 2n+6(\kappa _{s}^{+}+\kappa _{l}^{+}+\ell )+1%
\right] \qquad \text{for }n,\ell \in \mathbb{N}_{0}.
\end{equation}
Let us now see in a concrete case how the deformation weakens the
constraints and how it influences the physics of the models.

\subsubsection{The deformed case}

We may now consider various types of deformations (\ref{hj}) or (\ref{def2})
depending in addition on the possible selections for the deformation
functions $R(\varepsilon )$ and $I(\varepsilon )$. We consider the deformed $%
G_{2}$-Calogero model, with the deformation (\ref{hj}) and the simplest
choice for the deformation functions $R(\varepsilon )=\cosh \varepsilon $
and $I(\varepsilon )=\sqrt{3}\sinh \varepsilon $. This leads to the
differential equations (\ref{se1}) and (\ref{se2}) with a shifted $\phi
\rightarrow \phi +i\varepsilon $, i.e. the wavefunctions are simply obtained
from (\ref{phi1}), (\ref{phi2}) by $\eta _{\phi }\Psi (R,r,\phi )$ with $%
\eta _{\phi }=\exp (p_{\phi }\varepsilon )$. However, there is a small
change in the physical interpretation. From the discussion of the previous
subsection follows that $P1,P2$ and $P4$ still have to be implemented on
physical grounds, but to demand $P3$ lacks any justification, since the
wavefunctions are regularized and no longer diverge. Therefore $P3$ can be
relaxed. Consequently we end up with the modified energy spectrum 
\begin{equation}
E_{n\ell }^{\pm }=2|\omega |\left[ 2n+6(\kappa _{s}^{\pm }+\kappa _{l}^{\pm
}+\ell )+1\right] \qquad \text{for }n,\ell \in \mathbb{N}_{0},
\end{equation}
such that besides the energies $E_{n\ell }^{+}$ we may now also encounter
the energies $E_{n\ell }^{-}$. Note that we have a degeneracy $E_{n\ell
}^{+}=E_{n^{\prime }\ell ^{\prime }}^{-}$ whenever 
\begin{equation}
n^{\prime }-n+3(\ell ^{\prime }-\ell )=\frac{3}{2}\sqrt{1+4g_{s}}+\frac{3}{2}%
\sqrt{1+4g_{l}}.
\end{equation}
A similar observation was made for $A_{2}$-Calogero model in \cite{Milos}.

Alternatively we may investigate the deformed $G_{2}$-Calogero model (\ref%
{alte}) based on the deformed roots (\ref{def2}) with deformation $%
R(\varepsilon )=1$ and $I(\varepsilon )=\varepsilon /r$. The wavefunctions
are easy to construct in this case when we break the invariance under the
extended Coxeter group $\mathcal{W}^{\mathcal{PT}}$ by restricting the sum
in the potential to the positive or negative roots only and scaling the
coupling constants $g_{s},g_{l}$ by a factor of $2$. Then the corresponding
wavefunctions result from (\ref{phi1}), (\ref{phi2}) as $\eta _{r}^{\pm
}\Psi (R,r,\phi )$ with $\eta _{r}^{\pm }=\exp (\pm p_{r}\varepsilon )$. For
each of the models the constraints $P1,P3$ and $P4$ still hold on physical
grounds, but as the divergence at $r=0$ for $\chi (r)$ has vanished we no
longer have to demand $P2$. This means for both models, that is either
extending the roots just over the positive or just over the negative roots,
we have the identical energy spectra 
\begin{equation}
E_{n}^{\pm }=2\left| \omega \right| (2n\pm \lambda +1),
\end{equation}
thus allowing in addition to $E_{n}^{+}$ also $E_{n}^{-}$. We encounter the
degeneracy $E_{n}^{+}=$ $E_{n^{\prime }}^{-}$ when $\lambda =n^{\prime }-n$.
Due to the identity

\begin{equation}
z^{m-n}\Gamma (n+1)L_{n}^{m-n}\left( z^{2}\right) =(-z)^{n-m}\Gamma
(m+1)L_{m}^{n-m}\left( z^{2}\right)
\end{equation}
we find in that situation the wavefunction are related as 
\begin{equation}
\chi _{n}^{\lambda }(r+i\varepsilon )=(-1)^{n^{\prime }-n}\chi _{n^{\prime
}}^{-\lambda }(r+i\varepsilon ).
\end{equation}
In general, we have the symmetry $\chi _{n}^{\lambda }(r)=(-1)^{\lambda
}\chi _{n}^{\lambda }(-r)$, such that we can related the wavefunctions of
the positive root model $\chi _{n,\text{pos}}^{\lambda }(r)$ and the
negative root model $\chi _{n,\text{neg}}^{\lambda }(r)$ by an anyonic
statistic as $\chi _{n,\text{pos}}^{\lambda }(r)=(-1)^{\lambda }\chi _{n,%
\text{neg}}^{\lambda }(r)$.

\section{Conclusions}

We have demonstrated that the Coxeter group represented in $\mathbb{R}^{n}$
can be deformed in a systematic way to the $\mathcal{PT}$-symmetrically
extended Coxeter group $\mathcal{W}^{\mathcal{PT}}$ represented in $\mathbb{R%
}^{n}\oplus i\mathbb{R}^{n}$. As we have shown there are various ways to
achieve this. We may deform the roots across the hyperplanes through the
origin orthogonal to each root either by taking the imaginary part to be a
vector in this hyperplane (\ref{hj}) or a vector orthogonal to it (\ref{alt}%
). As a natural application one may seek for models for which this group
constitutes a symmetry group. CMS-models are well known to be invariant
under the action of $\mathcal{W}$ and we have demonstrated how they may be
deformed such that they remain invariant under the action of $\mathcal{W}^{%
\mathcal{PT}}$. In fact one simply needs to replace the roots $\alpha _{i}$
by their deformed counterparts $\tilde{\alpha}_{i}$. We have worked out the $%
A_{2}$ and $G_{2}$-cases in some detail by constructing explicitly the
deformed root systems and applying them thereafter to the corresponding
CMS-models. When specializing the deformation functions $R(\varepsilon )$
and $I(\varepsilon )$ in a certain way some easy cases resemble the
undeformed case with some simple shifts when transformed to Jacobian
relative coordinates, which allowed to determine their corresponding
eigensystems. We discussed that as a consequence of the deformation the
physical reason leading to some constraints vanishes, such that various
restrictions on the parameter space of the model may be relaxed.

Various open challenges remain, as for instance to establish whether the
deformations studied here preserve integrability, analogously to what has
been established in \cite{AF} for the different types of deformation, to
investigate models for different choices for the functions $R(\varepsilon )$
and $I(\varepsilon )$ and to study in detail Coxeter groups of higher rank,
together with their applications, such as the CMS-models \cite{prep}. Models
with different choices for the deformation functions will certainly also
lead to non-Hermitian Hamiltonians with real spectra, which may be explained
by the built in $\mathcal{PT}$-symmetry \cite{Bender:2002vv,EW} or
pseudo-Hermiticity \cite{Mostafazadeh:2002hb}.

\medskip \noindent \textbf{Acknowledgments:} AF would like to thank Paulo Gon%
\c{c}alves de Assis for discussions. The participation of MZ supported by
the M\v{S}MT \textquotedblleft Doppler Institute\textquotedblright project
Nr. LC06002, by GA CR grant Nr. 101/1307 and by the Institutional Research
Plan AV0Z10480505.


\end{document}